\documentclass[prb,twocolumn,showpacs,superscriptaddress]{revtex4-1}

\usepackage{graphicx}
\def \duud {\downarrow \uparrow \uparrow \downarrow}

\begin{document}
\title{The magnetic and crystal structures of $\rm Sr_2IrO_4$:
A neutron diffraction study}

\author{Feng Ye}
\affiliation{Quantum Condensed Matter Division, Oak Ridge National Laboratory,
Oak Ridge, Tennessee 37831, USA}
\affiliation{Center for Advanced Materials, Department of Physics and
Astronomy, University of Kentucky, Lexington, Kentucky 40506, USA}
\author{Songxue Chi}
\author{Bryan~C.~Chakoumakos}
\affiliation{Quantum Condensed Matter Division, Oak Ridge National Laboratory,
Oak Ridge, Tennessee 37831, USA}
\author{Jaime~A.~Fernandez-Baca}
\affiliation{Quantum Condensed Matter Division, Oak Ridge National Laboratory,
Oak Ridge, Tennessee 37831, USA}
\affiliation{Department of Physics and Astronomy, University of Tennessee,
Knoxville, Tennessee 37996, USA}
\author{Tongfei Qi}
\author{G.~Cao}
\affiliation{Center for Advanced Materials, Department of Physics and
Astronomy, University of Kentucky, Lexington, Kentucky 40506, USA}
\date{\today}

\begin{abstract}
We report a single-crystal neutron diffraction study of the layered $\rm
Sr_2IrO_4$. This work unambiguously determines the magnetic structure of the
system and reveals that the spin orientation rigidly tracks the staggered
rotation of the $\rm IrO_6$ octahedra in $\rm Sr_2IrO_4$.  The long-range
antiferromagnetic order has a canted spin configuration with an ordered moment
of 0.208(3) $\mu_B$/Ir site within the basal plane;  a detailed examination of
the spin canting yields 0.202(3) and 0.049(2) $\mu_B$/site for the $a$ axis
and the $b$ axis, respectively. It is intriguing that forbidden nuclear
reflections of space group $I4_1/acd$ are also observed in a wide temperature
range from 4~K to 600~K, which suggests a reduced crystal structure symmetry.
This neutron-scattering work provides a direct, well-refined experimental
characterization of the magnetic and crystal structures that are crucial to
the understanding of the unconventional magnetism exhibited in this unusual
magnetic insulator.
\end{abstract}

\maketitle

The 5$d$-based iridates have continuously provided a fertile playground for
the studies of novel physics driven by the spin-orbit interaction (SOI). It is
believed that SOI (0.4 - 1 eV), which is proportional to $Z^4$ ($Z$ is the
atomic number), plays a critical role in the iridates, and rigorously competes
with other relevant energies, particularly the on-site Coulomb interaction $U$
($0.4 - 2.5$ eV), which is significantly reduced because of the extended
nature of the 5$d$ orbitals.  A new balance between the competing energies is
therefore established in the iridates and drives exotic quantum phases.
Recent experimental observations and theoretical proposals for the iridates
have captured the intriguing physics driven by SOI and examples include the
following: the $\rm J_{eff} = 1/2$ Mott state,\cite{Kim08,Moon09,Kim09,Ge11}
superconductivity,\cite{Wang11,You12} a correlated topological insulator with
large gaps,\cite{Shitade09,Kim12e} spin liquid in a hyperkagome
structure,\cite{Okamoto07} Weyl semimetal with Fermi arcs,\cite{Wan11} the
Kitaev mode,\cite{Jackeli09,Singh12} and three-dimensional (3D) spin liquid
with Fermionic spinons.\cite{Zhou08}

Among all the iridates studied, the single layer $\rm Sr_2IrO_4$ has been
subjected to the most extensive investigations due to its structural and
electronic similarities to the undoped high-$\rm T_C$ cuprates such as $\rm
La_2CuO_4$. This magnetic insulator was proposed to be an effective $\rm
J_{eff}=1/2$ Mott-Hubbard state arising from the SOI.\cite{Kim08,Kim09}
Although the insulating ground state has been established by angle-resolved
photoemission spectroscopy\cite{Kim08} and resonant x-ray scattering (RXS)
measurements,\cite{Kim09} some critical insights into the crystal and magnetic
structures remain conspicuously elusive. For example, the strong SOI limit
$\rm J_{eff}=1/2$ ground-state scenario has been recently challenged by 
x-ray absorption spectroscopy,\cite{Haskel12} time-resolved optical
studies,\cite{Hsieh12} and theory.\cite{Chapon11} The nature of the weak
ferromagnetism arising from the canted antiferromagnetic (AFM) order is not
fully characterized experimentally. This is primarily due to the lack of large
single crystals and the strong absorbing cross section of the Ir ions that
prevent a comprehensive neutron study.  Here we report the results of a
neutron diffraction investigation of single-crystal $\rm Sr_2IrO_4$. The
central findings of this work are the following: (1) The magnetic and crystal
structures are completely determined; (2) the system undergoes an
antiferromagnetic transition at 224(2)~K with an ordered moment of
0.208(3)~$\mu_B$/Ir site and a canted spin configuration within the basal
plane; and (3) the spin orientation is intimately associated with the rotation
of the $\rm IrO_6$ octahedra, which results in 0.202(3) and 0.049(2)
$\mu_B$/Ir site for the $a$ axis and the $b$ axis, respectively.  In addition,
nuclear reflections incompatible with the previously reported space group (SG)
are observed and indicate a possible lowering of the structural symmetry.

The $\rm Sr_2IrO_4$ single crystal studied ($2\times2\times1$mm$^3$,
mass=8~mg) was grown using self-flux techniques.\cite{Cao98} Because the
iridium is highly neutron absorbing, the equal-dimensional shaped crystal
simplifies the necessary absorption correction.\cite{abscor} The neutron diffraction
measurements were carried out at the HB1A, HB1 triple axis spectrometers, and
the HB3A four circle diffractometer at the High Flux Isotope Reactor at the
Oak Ridge National Laboratory. For the measurements using triple axis
spectrometers, the crystal was aligned in the $(h,0,l)$, $(h,h,l)$, $(0,k,l)$
and other scattering planes to probe various magnetic reflections. A
close-cycle refrigerator and high temperature furnace were employed to monitor
the $T$ dependence of the magnetic and nuclear reflections.

\begin{figure}[ht!]
\includegraphics[width=3.3in]{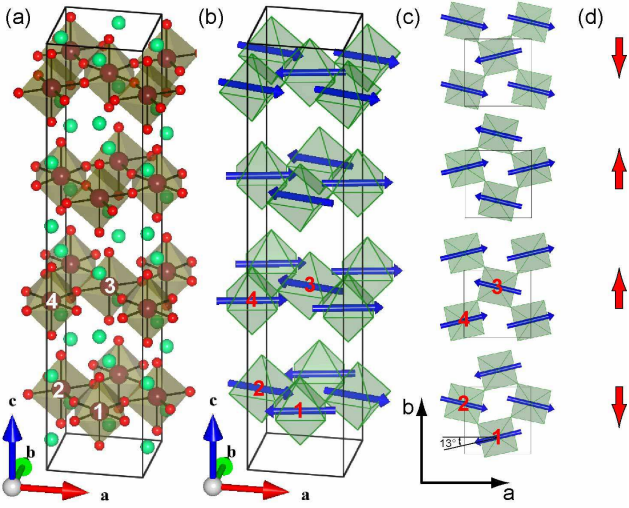}
\caption{(Color online) (a) The crystal structure of $\rm Sr_2IrO_4$ with
SG $I4_1/acd$ (setting 2). Each $\rm IrO_6$ octahedron rotates
~11.8$^\circ$ about the $c$ axis. The Ir atoms of the nonprimitive basis are
labeled 1, 2, 3, and 4 plus the body centering translation (1/2,1/2,1/2). (b)
The refined magnetic structure from single-crystal neutron diffraction
measurements.  (c) The same spin configuration projected on the basal planes.
(d) The net moment projected along the $b$ axis for individual layers.
}
\label{fig:structure}
\end{figure}

$\rm Sr_2IrO_4$ was reported to crystallize in a tetragonal structure (SG
$I4_1/acd$, No.~142) with $a=b=5.484\rm~\AA$ and $c=25.83\rm~\AA$ at 4~K.
\cite{Huang94,Crawford94} With reflection conditions compliant with the
$I4_1/acd$ symmetry, we have collected 137 nuclear reflections of $\rm
Sr_2IrO_4$ using HB3A for structure refinements. The most prominent features
of the crystal structure are the elongation of the $\rm IrO_6$ octahedra along
the $c$ axis (2.055~\AA~for the out-of-plane distance compared to
1.981~\AA~in-plane one), and the rotation of the octahedra with respect to the
$c$ axis about 11.8(1)$^\circ$ at 4~K. This leads to a
$\sqrt{2}\times\sqrt{2}$ expansion of unit cell in the basal plane compared to
the higher symmetry $\rm Sr_2RuO_4$ [Fig.~1(a)].

\begin{figure}[ht!]
\includegraphics[width=3.3in]{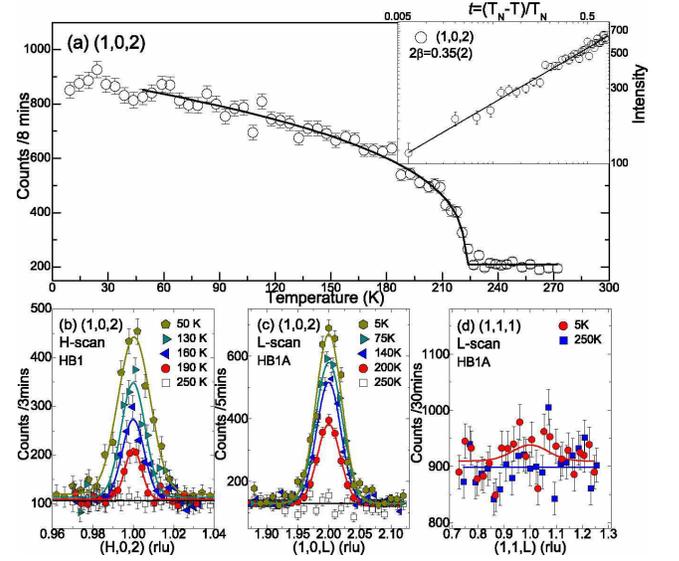}
\caption{(Color online) (a) The $T$-dependence of the magnetic (1,0,2)
reflection. Inset shows the intensity versus the reduced temperature
($t=|1-T/T_N|$) in logarithmic scale. The wave-vector scan along (b) the
[H,0,0] and (c) the [0,0,L] directions for the (1,0,2) peak at selected
temperatures that probe the in-plane and out-of-plane correlations.  (d)
Similar wave-vector scans for the magnetic (1,1,1) reflection above and below
$\rm T_N$. Note that the counting statistics is 10 times compared to those of
the strong reflection (1,0,2) peak.
}
\label{fig:OP102}
\end{figure}

The antitranslation in combination with the body centering dictates a (1,1,1)
magnetic propagation wave-vector, as discussed
previously.\cite{Chapon11,Lovesey12b} Figure \ref{fig:OP102}(a) displays the
$T$ dependence of the Bragg intensity ($I_B\propto |M_s|^2$, $M_s$ is the
order parameter) of the magnetic reflection (1,0,2). The intensity vanishes
around $T_N=224(2)$~K and is consistent with the magnetization
measurement.\cite{Cao98} Fitting the order parameter to the power-law scaling
function $I_B\approx|t|^{2\beta}$, where $t=1-T/T_N$ is the reduced
temperature, leads to the critical exponent $\beta=0.18(1)$. It apparently
deviates from the $\beta=0.325$ expected for a 3$D$ Heisenberg spin system.
Figures 2(b) and 2(c) illustrate the wave-vector scans within and perpendicular to
the basal plane at several temperatures. In both cases, the lineshape of the
magnetic scattering evolves into a Gaussian profile below $T_N$, signaling the
formation of the long-range magnetic order.  Our observation is in accord with the
RXS studies indicating that a short-range Heisenberg spin fluctuation occurs
only in a paramagnetic state.\cite{Fujiyama12}

\begin{table}[ht!]
\caption{Basis vectors (BVs) for the SG $I4_1/acd$ with magnetic propagation
vector ${\bf k}$=(1,1,1).  The decomposition of the magnetic representation is
$\Gamma_{\rm mag}=2\Gamma^2_1+2\Gamma^2_2+2\Gamma^1_3+2\Gamma^2_4$. The atoms
of the nonprimitive basis are located at
1:(1/2,1/4,1/8),2:(0,3/4,1/8),3:(1/2,3/4,3/8),4:(0,1/4,3/8) (Figure 1). For
clarity, only the in-plane BVs are listed.
}
\label{tab:BV}
\begin{ruledtabular}
\begin{tabular}{rrrrrrrrrrrr}
& & & \multicolumn{3}{r}{component} & & & &\multicolumn{3}{r}{component} \\ \cline{4-6} \cline{10-12}
IR & BV & atom           &$m_a$ &$m_b$ &$m_c$ & IR & BV & atom   &$m_a$
&$m_b$ &$m_c$ \\ 
    \hline
$\Gamma_1$ & $\psi_1$ & 1 & 1 & 0& 0& $\Gamma_2$ & $\psi_5$ &1 & 1 & 0& 0 \\
           &          & 2 & 1 & 0& 0&            &          &2 & 1 & 0& 0 \\
           &          & 3 & 1 & 0& 0&            &          &3 &-1 & 0& 0 \\
           &          & 4 & 1 & 0& 0&            &          &4 &-1 & 0& 0 \\
           & $\psi_2$ & 1 & 0 & 1& 0&            & $\psi_6$ &1 & 0 & 1& 0 \\
           &          & 2 & 0 & 1& 0&            &          &2 & 0 & 1& 0 \\
           &          & 3 & 0 &-1& 0&            &          &3 & 0 & 1& 0 \\
           &          & 4 & 0 &-1& 0&            &          &4 & 0 & 1& 0 \\
           & $\psi_3$ & 1 & 1 & 0& 0&            & $\psi_7$ &1 & 1 & 0& 0 \\
           &          & 2 &-1 & 0& 0&            &          &2 &-1 & 0& 0 \\
           &          & 3 & 1 & 0& 0&            &          &3 &-1 & 0& 0 \\
           &          & 4 &-1 & 0& 0&            &          &4 & 1 & 0& 0 \\
           & $\psi_4$ & 1 & 0 & 1& 0&            & $\psi_8$ &1 & 0 & 1& 0 \\
           &          & 2 & 0 &-1& 0&            &          &2 & 0 &-1& 0 \\
           &          & 3 & 0 &-1& 0&            &          &3 & 0 & 1& 0 \\
           &          & 4 & 0 & 1& 0&            &          &4 & 0 &-1& 0 \\
  \end{tabular}
\end{ruledtabular}
\end{table}

\begin{figure}[ht!]
\includegraphics[width=3.4in]{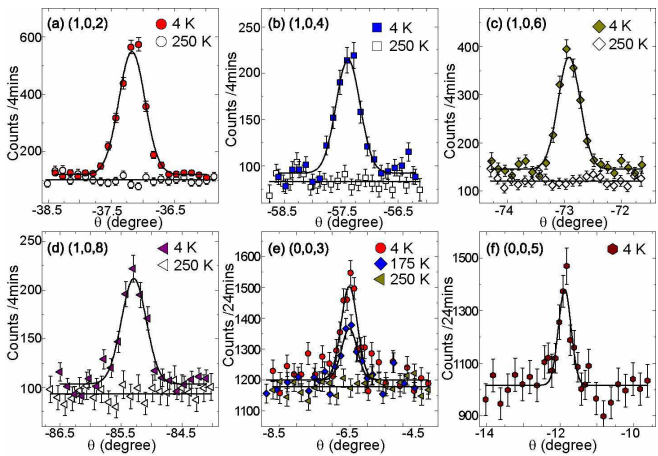}
\caption{(Color online) Selected rocking scans at 4~K and 250~K for the
(a) (1,0,2), (b) (1,0,4), (c) (1,0,6), (d) (1,0,8), (e) (0,0,3), and (f)
(0,0,5) magnetic reflections. The weaker $(0,0,2n+1)$ are measured with much
longer counting time.
}
\label{fig:rocking}
\end{figure}

A quantitative characterization of the magnetic structure and moment size of
the Ir$^{4+}$ ions can be obtained by a comprehensive survey of the magnetic
reflections in conjunction with the model calculation. Figure 3 shows the
neutron diffraction scans at selected reflections. The disappearance of the
scattering above $T_N$ and decrease in intensity at large momentum transfer
indicate their magnetic nature. Differing from the early RXS studies where the
magnetic reflections are present only at $(0,1,4n+2)$ and
$(1,0,4n)$,\cite{Kim09} our neutron diffraction shows additional Bragg peaks
at the $(0,1,4n)$ and $(1,0,4n+2)$ positions. The nearly identical intensity at
equivalent wave-vectors (1,0,2) and (0,1,2) indicates the crystal has equally
populated magnetic domains. Note that the structural refinement with the same
sample cannot determine whether the system is structurally
twinned.\cite{xray-twinning} The presence of both type of reflections strongly
suggests that they originate from the twinned crystallographic domains.
According to the Landau theory, the symmetry properties of the magnetic
structure can be described by only one irreducible representation (IR). With
Ir ions located at the $8a$ Wyckoff positions for the SG $I4_1/acd$ and the
propagation wave-vector $q_M=(1,1,1)$, the magnetic representation can be
decomposed into $\Gamma_{\rm
mag}=2\Gamma^2_1+2\Gamma^2_2+2\Gamma^1_3+2\Gamma^1_4$, where $\Gamma_1$,
$\Gamma_2$ are the two-dimensional IRs with basis vectors lying in the $ab$
plane, $\Gamma_3$, $\Gamma_4$ are the one-dimensional IRs with moments
pointing parallel the $c$-axis.  Since the magnetic susceptibility suggests
that the spin easy axis lies in the basal plane,\cite{Cao98} we exclude spin
configurations associated with $\Gamma_3$ and $\Gamma_4$ in the analysis.
Table~\ref{tab:BV} lists the basis vectors of IRs $\Gamma_1$ and $\Gamma_2$.
In particular, the spin structure based on linear combination of $\psi(2)$ and
$\psi(3)$ of $\Gamma_1$ has a $(+-+-)$ configuration along the $a$-axis (or
the M4 structure described in Refs.~\onlinecite{Chapon11,Lovesey12b}) and
$(++--)$ along the $b$-axis for the labeled Ir ions in Figure~1. In contrast,
the linear combination of $\psi(5)$ and $\psi(8)$ in $\Gamma_2$ gives $(++--)$
along the $a$-axis and $(+-+-)$ along the $b$-axis (the M3 configuration).
These spin structures derived from representation analysis using BasIreps
program \cite{Rodriguezcarvajal93} are in accord with the results from
previous neutron powder diffraction.\cite{Lovesey12b}

\begin{table}[ht!]
\caption{ Comparison of observed and calculated magnetic intensities from
two symmetry compatible spin models. To get the scale factor, separate sets of
nuclear reflections were collected at HB3A with incident neutron wavelength of
1.5424 and 1.003 $\AA$, respectively. Additional 37 nuclear reflections were
collected at HB1A for intensity normalization.\cite{2axisnote}}
\label{tab:comp}
\begin{ruledtabular}
\begin{tabular}{lrrr}
reflection & observation &  M4 model & M3 model  \\
    \hline
(0,0,3) & $0.26\pm0.03$ & 0.25	& 0.25    \\
(0,0,5) & $0.20\pm0.03$ & 0.20 	& 0.20    \\
(1,1,1) & $0.08\pm0.07$	& 0.08 	& 0.08    \\
(0,1,2) & $6.80\pm0.17$ & 6.99 	& 1.05    \\
(0,1,6) & $4.72\pm0.32$ & 4.50	& 2.73    \\
(1,0,2) & $6.99\pm0.18$ & 6.99	& 1.05    \\
(1,0,4) & $2.33\pm0.22$ & 2.48 	& 5.81    \\
(1,0,6) & $4.72\pm0.32$ & 4.51 	& 2.73    \\
(1,0,8) & $2.56\pm0.32$ & 2.28 	& 3.01    \\
(1,0,14)& $0.88\pm0.21$ & 0.54 	& 0.48    \\
(1,0,16)& $0.18\pm0.09$ & 0.24 	& 0.26    \\
(1,2,0) & $1.53\pm0.36$ & 1.76 	& 0.43    \\
(1,2,4) & $1.14\pm0.23$ & 1.46	& 0.52    \\
(1,2,8) & $0.85\pm0.12$ & 0.82	& 0.45    \\
  \end{tabular}
\end{ruledtabular}
\end{table}

Table II compares the expected intensities for the two relevant spin models and
the experimental observations. The M4 and M3 configurations each have distinct
distributions of magnetic scattering intensities.\cite{baxis} For example, the
collinear structure with $a$-axis $(+-+-)$ components produces the strongest
scattering at the (0,1,2) reflection and gives zero intensity at the (1,0,0)
Bragg point.  However, the $(++--)$ collinear state associated with the M3
configuration will generate the strongest scattering at the (1,0,0) peak, which
is not observed experimentally.  The neutron diffraction results shown in
Table II clearly support the M4 spin configuration and confirm the previous
neutron diffraction work on the powder sample.\cite{Lovesey12b} To test
whether there are additional canted moments along $b$ axis with the $(++--)$
configuration within $\Gamma_1$, we probed the scattering at the expected
$(0,0,2n+1)$ reflections.  Figures 3(d) and 3(e) display the scans at the (0,0,3)
and (0,0,5) Bragg peaks.  Although much weaker, the magnetic scattering is
clearly present at low $T$ and confirms the staggered AFM order propagating
along the $c$ axis. A total of 14 magnetic reflections combined with 137
nuclear reflections allow an accurate determination of the spin structure and
the associated moment.  Using the M4 spin model and the magnetic form factor
for Ir$^{4+}$,\cite{Kobayashi11} we have obtained $m_a=0.202(3)~\mu_B$ along
the $a$ axis and $m_b=0.048(2)~\mu_B$ along the $b$ axis, yielding a total
moment of $0.208(3)~\mu_B$/Ir$^{4+}$ site. This value is smaller than
0.36(6)~$\mu_B$ from a recent single crystal neutron-scattering
study\cite{Dhital12b} but quite consistent with the powder neutron diffraction
results in which the upper limit of the moment does not exceed
0.29(4)~$\mu_B$.\cite{Lovesey12b} The magnetic configuration in
Figs.~1(b)-1(d) show that spins projected along the $b$-axis have a staggered
$\duud$ pattern along the $c$ axis, with Ir spins deviating $13(1)^\circ$ away
from the $a$ axis [see Fig.~1(d)]. This spin canting rigidly tracks the
staggered octahedral rotation, as illustrated in a previous RXS
study.\cite{Kim08} This remarkable correlation proves the existence of strong
magnetoelastic coupling in the iridate, which is also suggested in
experimental studies of transport and magnetic properties of the
system.\cite{Ge11,Chikara09}

Theoretically, the spin Hamiltonian in the strong SOI limit includes the
isotropic coupling ($J$) and the Dzyaloshinsky-Moriya interaction ($D$) caused
by the lattice distortion.\cite{Jackeli09} The spin canting is governed by the
ratio of $D$ and $J$ and is solely determined by the lattice distortion. This
explains the relatively large spin canting in the $5d$ system compared to that
in $\rm La_2CuO_4$ where SOI is insignificant (SOI $\propto Z^4$, $Z$=29 and 77
for Cu and Ir, respectively). The measured magnetic moment is much
smaller than $1~\mu_B$ conventionally anticipated for a $\rm S=1/2$ system
but is similar to those of other iridates, such as $\rm Na_2IrO_3$ and $\rm
BaIrO_3$ where the saturated moment is less than 20\% of $\rm 1~\mu_B$/Ir
\cite{Cao98,Ye12}. The significantly reduced moment might be attributed to the
strong hybridization of the Ir $5d$ orbital with the ligand oxygen $2p$
orbital because of the large spatial extend of $5d$ wave functions 
or the axial distortion of $\rm IrO_6$ octahedra away from the cubic
symmetry.\cite{Chapon11,Lovesey12b} Although the latter has been invoked to
explain the reduced moment, it is inconsistent with the branching ratio (BR)
obtained from the x-ray absorption spectrum.\cite{Clancy12,Laan88} The reduced
$\langle {\bf S \cdot \bf L} \rangle$ caused by the decreased moment makes the
corresponding BR values far too small compared to the measured one.  It was
argued that the moment value and BR are irreconcilable using only one $t_{2g}$
electron and $j=5/2$. For instance, Laguna-Marco {\it et al.}~have shown in a
multielectron simulation in $\rm BaIrO_3$ that $\rm J_{eff}=1/2$ accounts for
only half of $\langle {\bf S \cdot \bf L} \rangle$ required in BR to match the
experimental determined value, while the remaining half is induced by
spin-orbit mixing of the $t_{2g}$ and $e_g$ states.\cite{Laguna-Marco10}

\begin{figure}[ht!]
\includegraphics[width=3.3in]{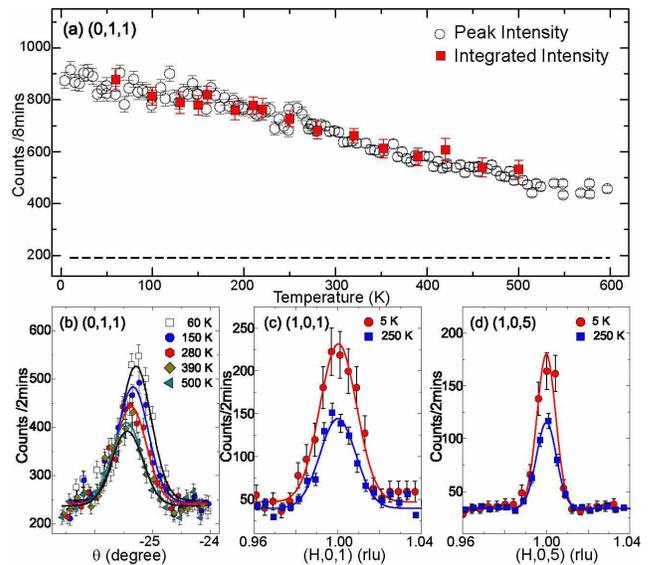}
\caption{(Color online)
(a) The $T$-dependence of the structural (0,1,1) reflection.  Open circles are
the peak intensity and solid squares the integrated intensity. Dashed line is
the background derived from the Gaussian fit to the rocking scan. (b) The
rocking scans of the (0,1,1) peak at $T$= 60, 150, 280, 390, and 500~K.  The
wave-vector scans along the [1,0,0] direction for (c) the (1,0,1) and (d) the
(1,0,5) reflections at 5~K and 250~K.
}
\label{fig:OP101}
\end{figure}

The observation of the magnetic $(0,1,4n+2)$ and $(1,0,4n)$ indicates either
the breakdown of the tetragonal symmetry of the system or a weak coupling
between the magnetic and lattice degrees of freedom. To understand the
possible structural origin of the anomalous magnetic behavior, we have
surveyed extensively in reciprocal space and observed the presence of nuclear
reflections $(odd~h,0,odd~l)$ that are not allowed in SG $I4_1/acd$.  Similar
behavior is also observed in a recent single crystal neutron diffraction
work.\cite{Dhital12b} Figure 4(b) shows the rocking scans of the (0,1,1)
reflection at selected temperatures. The intensity continuously decreases
on warming and shows no sign of transition to 600~K. The reduction in
intensity cannot be accounted by the thermal vibration of the elements
(Debye-Waller factor). The lack of anomaly near $T_N$ is also consistent with
the transport,\cite{Cao98,Kini06} thermodynamic,\cite{Chikara09}, and optical
conductivity studies.\cite{Moon09,Hsieh12} Scans across other Bragg peaks of
(1,0,1) and (1,0,5) display similar violation of the required $(h=2n,0,l=2n)$
reflection condition. Although it cannot be completely ruled out that the
forbidden peaks might be due to the structural defects such as oxygen
vacancies commonly observed in oxides, the systematically enhanced intensities
of these forbidden peaks with isovalent Rh doping \cite{Ye13} suggest it is an
intrinsic property. If the observed forbidden peaks arise from the reduced
crystal symmetry, they would lead to possible nonisomorphic subgroups of
either $I4_1/a22$ (No.~98) or $I4_1/a$ (No.~88) due to the absent $c$- and
$d$-glide planes. The absence of scattering across the (1,1,0) reflection
further rules out the SG of $I4_1/a22$.  Such observation of reduced
structural symmetry that persists at a much higher temperature than $T_N$,
implies the formation of a crystallographic template for the low-$T$ spin
structure that changes the tetragonal symmetry.  This observation is certainly
intriguing and the origin of it remains to be understood.

It is established that the magnetic and electronic properties are highly
susceptible to slight impurity doping for Sr, Ir, or
oxygen.\cite{Ge11,Chikara09,Calder12,Qi12,Korneta10} For example, doping Mn
results in an spin-flop transition with moments aligning along the
$c$-axis.\cite{Calder12} The remaining $\rm J_{eff}=1/2$ state revealed by RXS
measurement suggests its robustness against the alternation of spin structure.
On the other hand, replacing Ir with isovalent Rh$^{4+}$ leads to a rich phase
diagram of metal-insulator transition tuned by SOI.\cite{Qi12} The transition
was explained by the effective reduction of the splitting between the $\rm
J_{eff}=1/2$ and $\rm J_{eff}=3/2$ bands due to the reduced SOI; this in turn
alters the relative strength of the SOI and the crystal electric field (CEF)
that dictates the magnetic state.  This notion is also consistent with a recent
theoretical proposal that the change of CEF associated with the underlying
structure could be critical to determine the magnetic ground states. The
present single-crystal neutron diffraction unambiguously determines the
magnetic structure and proves the rigid coupling of the spin canting with the
rotation of the $\rm IrO_6$ octahedra.  These findings finally fill the
longstanding gap in our understanding of the magnetic properties in $\rm
Sr_2IrO_4$, an archetype of the $\rm J_{eff}=1/2$ insulators.

We thank Q.~Huang, S.~Lovesey, D.~Khalyavin and G.~Khaliullin for invaluable
discussions.  Research at ORNL's High Flux Isotope Reactor was sponsored by
the Scientific User Facilities Division, Office of Basic Energy Sciences, U.S.
Department of Energy.  The work at University of Kentucky was supported by NSF
through Grants No.~DMR-0856234 and EPS-0814194.


\begin{thebibliography}{10}
\bibitem{abscor}
The absorption corection factor has been numerically applied to the collected
reflections based on the sample shape. The mean transmission ($T$) is 0.92 with the
minimum at the (0,0,8) reflection ($T=0.79$) and the maximum at the (2,1,1)
peak ($T=0.95$).
\bibitem{baxis}
Because of the tetragonal symmetry of the crystal structure, the linear
combination of $\psi(1)$ and $\psi(4)$ within the same $\Gamma_1$ IR can also
describe the observed intensities. For simplicity, we use $\psi(2)$ and
$\psi(3)$ to be consistent with the spin configuration reported in
Ref.~\onlinecite{Lovesey12b}.
\end{thebibliography}
\end{document}